\documentclass[aps,prd,twocolumn,showpacs]{revtex4} 

\usepackage{graphicx}

\begin{document}

\title{Quantum Coherence of Relic Neutrinos}
\author{George M. Fuller and Chad T. Kishimoto}
\affiliation{Department of Physics, University of California, San Diego, La Jolla, CA
92093-0319}

\date{\today}

\begin{abstract}
We argue that in at least a portion of the history of the universe the relic background neutrinos are spatially-extended, coherent superpositions of mass states. We show that an appropriate quantum mechanical treatment affects the neutrino mass values derived from cosmological data.  The coherence scale of these neutrino flavor wavepackets can be an appreciable fraction of the causal horizon size, raising the possibility of spacetime curvature-induced decoherence.
\end{abstract}


\maketitle

In this letter we point out a curious feature of the neutrino background and the potential implications of this both for cosmology and for neutrino physics.  Neutrinos and antineutrinos in the early universe should be in thermal and chemical equilibrium with the photon- and $e^\pm$-plasma for temperatures $T > T_{\rm weak} \sim 1\,{\rm MeV}$. For $T\ll T_{\rm weak}$, the neutrinos and antineutrinos will be completely decoupled, comprising seas of ``relic'' particles freely falling through spacetime with energy-momentum and flavor distributions reflecting pre-decoupling equilibrium plus the expansion of the universe. This is analogous to the presently-decoupled cosmic microwave background (CMB) photons. These photons had been coupled and in thermal equilibrium in the early universe when $T> T_{\rm em} \approx 0.26\,{\rm eV}$. 

Experiments have demonstrated that the neutrino energy (mass) eigenstates $\vert\nu_i\rangle$ are not coincident with the weak interaction (flavor)
eigenstates $\vert \nu_\alpha\rangle$. These bases are related by the Maki-Nakagawa-Sakata (MNS) matrix, ${\vert \nu_\alpha\rangle =\sum_{i}{U_{\alpha i}^* \vert\nu_i\rangle}}$, where $\alpha=e,\mu,\tau$, and $i=1,2,3$ denotes the mass eigenstates, with corresponding vacuum mass eigenvalues $m_i$, and $U_{\alpha i}$ are the unitary transformation matrix elements (parameterized by 3 mixing angles $\theta_{1 2}$, $\theta_{2 3}$, $\theta_{1 3}$, and a $CP$-violating phase $\delta$). 

\begin{figure}
\includegraphics[width=1.2in, angle=90]{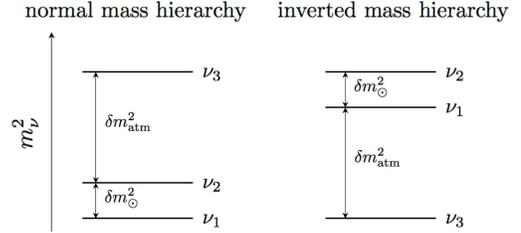}
\caption{Neutrino mass eigenvalue ordering in the ``normal mass hierarchy'' at left; ``inverted mass hierarchy'' at right.}
\label{figure1}
\end{figure}

Absolute neutrino masses remain unknown, but the neutrino mass-squared differences are measured. Studies of atmospheric neutrinos reveal a characteristic mass-squared splitting, $\vert\delta m^2_{\rm atm}\vert \approx 2.4\times{10}^{-3}\,{\rm eV}^2$, associated with $\nu_\mu\rightleftharpoons\nu_\tau$ mixing with $\sin^2 \theta_{ 2 3} \approx 0.50$. Likewise, solar neutrino observations and reactor-based experimental data suggest that the solar neutrino deficit problem is solved by flavor transformation in the $\nu_e\rightleftharpoons\nu_{\mu/\tau}$ channel and that the characteristic mass-squared difference for this solution is $\delta m^2_\odot \approx {7.6\times{10}^{-5}\,{\rm eV}^2}$, and that $\sin^2\theta_{12} \approx 0.31.$  Current experimental limits on $\theta_{13}$ are $\sin^2 \theta_{13} \leq 0.040$ ($2\sigma$ limit) \cite{stv08}. The $CP$-violating phase and ordering of the neutrino mass splittings remain unconstrained by experiment.  In the ``normal'' neutrino mass hierarchy the solar neutrino mass-squared doublet lies below the atmospheric doublet; in the ``inverted'' mass hierarchy it is the other way around. This is illustrated in Fig.\ \ref{figure1}.

For $T>T_{\rm weak}$, prior to decoupling, solutions of the quantum kinetic equations for neutrino flavor evolution show that neutrinos will be forced by weak interaction-mediated scattering into flavor eigenstates \cite{kf08, bvw99, dhpprs02}.  At epoch $T_{\rm weak}$, corresponding to scale factor $a_{\rm weak}$, the neutrino distribution functions for each flavor will be two-parameter, Fermi-Dirac black bodies, with the number density of $\nu_\alpha$'s in energy interval $dE_\nu$ given by
\begin{equation} 
\label{dist}
dn_{\nu_\alpha}= {{1}\over{2\pi^2}}\cdot {{p^2}\over{ e^{E_\nu/T_{\rm weak} - \eta_{\nu_\alpha}} +1}} \left( \frac{d E_\nu}{d p} \right)^{-1} \,d E_\nu,
\end{equation} 
where we employ natural units with $\hbar = c = k_B = 1$, the degeneracy parameter (ratio of chemical potential to temperature) for neutrino species $\nu_\alpha$ is $\eta_{\nu_\alpha}$, and where the neutrino energy-momentum dispersion relation is 
\begin{equation}
E_\nu = \sum_i \vert U_{\alpha i} \vert^2 \left( p^2 + m_i^2 \right)^{1/2},
\end{equation}
with $p = \vert {\bf p} \vert$ the magnitude of the spacelike momentum.  Here we will ignore small spectral distortions between $\nu_e$ and $\nu_{\mu, \tau}$ from $e^\pm$-annihilation.

Subsequent to decoupling, these pure flavor states can be regarded as collisionless, simply free falling through spacetime with their momentum components redshifting with the scale factor, $p \propto a^{-1}$.   As a result, the number density of $\nu_\alpha$ at an epoch with scale factor $a$ is
\begin{equation}
\label{timedist}
dn_{\nu_\alpha} (a) \approx \frac{T^3_\nu (a)}{2 \pi^2} \cdot \frac{\epsilon^2 d \epsilon}{e^{E_\nu(a) / T_{\rm weak} - \eta_{\nu_\alpha}} + 1},
\end{equation}
where $T_\nu (a) = T_{\rm weak} a_{\rm weak} / a$ is an effective ``temperature'' of the neutrinos, $\epsilon = p / T_\nu (a)$ is a co-moving invariant, and the energy-momentum dispersion relation is
\begin{equation}
\label{emomdisp}
E_\nu (a) \approx T_{\rm weak} \left( \epsilon^2 + \frac{\sum_i \vert U_{\alpha i} \vert^2 m_i^2}{T_{\rm weak}^2} \right)^{1/2} .
\end{equation}
In both of these expressions, we have neglected corrections of order $\left( m_i / T_{\rm weak} \epsilon \right)^4$ which are small since experimental measurements of neutrino masses have $m_i \lesssim 1 ~\rm{eV}$, $T_{\rm weak} \sim 1 ~\rm{MeV}$, and small values of $\epsilon$ are suppressed in the neutrino distribution function, Eq.\ (\ref{timedist}) ({\emph i.e.}, $\left( m_i / T_{\rm weak} \epsilon \right)^4 \lesssim 10^{-24}$).  The energy-momentum dispersion relation, Eq.\ (\ref{emomdisp}), leads to the standard choice for the effective mass of a neutrino in flavor state $\nu_\alpha$,
\begin{equation}
m^2_{\rm{eff}, \nu_\alpha} = \sum_i \vert U_{\alpha i} \vert^2 m_i^2.
\label{meff}
\end{equation}
This is the dynamical mass for an ultrarelativistic neutrino of flavor $\nu_\alpha$.  However, we will show that when the neutrino momentum redshifts to a point where the neutrino's kinematics become less relativistic \cite{ag97}, this effective mass is no longer relevant in characterizing the energy-momentum dispersion relation.

Examining Eqs.\ (\ref{timedist}) and (\ref{emomdisp}), we notice that the distribution function for $\nu_\alpha$ maintains a self-similar form with $\epsilon$ and $\eta_{\nu_\alpha}$ being co-moving invariants.  This form is suggestive of, though subtly different from a Fermi-Dirac blackbody with temperature $T_\nu$ and degeneracy parameter $\eta_{\nu_\alpha}$.  The energy distribution functions for neutrinos in mass eigenstates are a weighted sum of the flavor eigenstates,
\begin{equation}
d n_{\nu_i} = \sum_\alpha \vert U_{\alpha i} \vert^2 d n_{\nu_\alpha} .
\label{massdist}
\end{equation} 
If the degeneracy parameters for the three flavors were all equal, then the distribution function for the mass states would have the same Fermi-Dirac form.  This follows from unitarity, $\sum_\alpha \vert U_{\alpha i} \vert^2 = 1$.  However, in general the distribution functions of neutrinos in mass eigenstates would not have a Fermi-Dirac form where the degeneracy parameters were not identical for all three active flavors.

CMB- and large scale structure-derived neutrino mass limits typically are predicated on an assumption that these distribution functions have a Fermi-Dirac black body form. This assumption is invalid when the energy spectra of the various neutrino flavors are not identical, {\it i.e.,} when the neutrino degeneracy parameters (or, equivalently, the corresponding lepton numbers) are unequal. 

The measured solar neutrino mixing parameters coupled with Big Bang Nucleosynthesis (BBN) considerations limit $\eta_{\nu_\alpha} < 0.15$ \cite{abfw, kssw01} for all flavors and dictate that the degeneracy parameters be within an order of magnitude of each other \cite{abb02, abfw, dhpprs02}. Consequently, the spectral and number density differences for the neutrino flavors stemming from disparate lepton numbers cannot be large, but conceivably could be $ \lesssim 15\%$. Nevertheless, as neutrino mass limits improve with future higher precision CMB experiments and better large scale structure observations, it may be necessary to include the possibility of non-Fermi-Dirac energy spectra in these neutrino mass analyses.  For the remainder of this letter, we will assume that all the neutrino degeneracy parameters are zero to illustrate another important result.

\begin{figure}
\includegraphics[width = 2in, angle = 270]{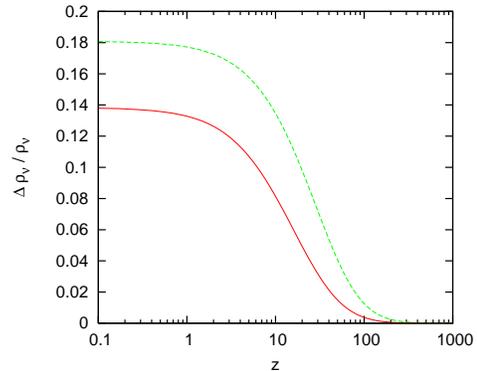}
\caption{\label{fig:omeganu} $(\rho_\nu^{(0)} - \rho_\nu) / \rho_\nu$ as a function of redshift.  Solid curve is for $m_1 = 1 ~\rm{meV}$ in the normal neutrino mass hierarchy and the dashed curve is for $m_3 = 1 ~\rm{meV}$ in the inverted hierarchy.}
\end{figure}

There is a more general issue associated with determining neutrino masses from cosmological data.  A naive method of calculating $\rho_\nu$, the energy density of neutrinos in the universe, is to use the effective mass in Eq.\ (\ref{meff}) and the number density distributions of neutrinos in flavor eigenstates,
\begin{equation}
\rho_\nu^{(0)} = \sum_\alpha \int ( p^2 + m^2_{\rm{eff}, \nu_\alpha} )^{1/2} d n_{\nu_\alpha} .
\label{eq:rhonu_naive}
\end{equation}
However, if we wish to calculate the energy density of these quantum mechanical particles, we ought to consider the energy (mass) eigenstates in the calculation,
\begin{eqnarray}
\rho_\nu & = & \sum_i \int ( p^2 + m_i^2 )^{1/2} d n_{\nu_i} \nonumber \\
 & = & \sum_{i, \alpha} \int \vert U_{\alpha i} \vert^2 ( p^2 + m_i^2 )^{1/2} d n_{\nu_\alpha} .
\end{eqnarray}
At late times (large scale factors or equivalently small redshifts), the momenta of the neutrinos have redshifted to a point where they are comparable to the neutrino masses.  In this case the quantum mechanically consistent calculation diverges from the naive effective mass approach.  Fig.\ \ref{fig:omeganu} shows the fractional difference between these two calculation techniques.  At large redshifts, neutrino momenta are large enough that all three mass eigenstates are ultrarelativstic and masses have little effect on the overall energy density.  At low redshifts, the kinematics of the neutrinos is no longer ultrarelativistic and rest mass becomes significant in the overall energy density.  At this point the difference in the two methods manifests itself in the difference between the quantum mechanical expectation value for the mass, $\langle m \rangle$, and the root-mean-squared mass expectation value, $\langle m^2 \rangle^{1/2}$.

As can be seen in Fig.\ \ref{fig:omeganu}, the disparity in calculating the value of $\rho_\nu$ becomes significant, of order $10 \%$, for redshifts of order $1 - 10$, at least for $m_1 \sim 1 ~\rm{meV}$.  This could be important because this is the epoch of structure formation in the universe where the energy density in neutrinos and the character of their kinematics are relevant to the formation of large scale structures.  

One method of experimentally determining the mass of the neutrino using the CMB is by inferring the transfer function in the matter power spectrum at large wavenumbers (or small scales).  Massive neutrinos contribute to the closure fraction in cold dark matter at the current epoch (as inferred by measurements of the CMB power spectrum), but at higher redshifts these neutrinos may act as hot dark matter.  As a result, the amount of cold dark matter at these earlier epochs is reduced, suppressing the formation of large scale structure compared to a situation without massive neutrinos.  A comparison of the observationally-inferred matter power spectrum with the power spectrum expected without the effects of massive neutrinos implies $\Omega_\nu$, the closure fraction contributed by neutrinos.  Determining the energy density in neutrinos at late times is an important aspect of this procedure and is necessary in deriving neutrino masses.  

\begin{figure}
\includegraphics[width=2in, angle = 270]{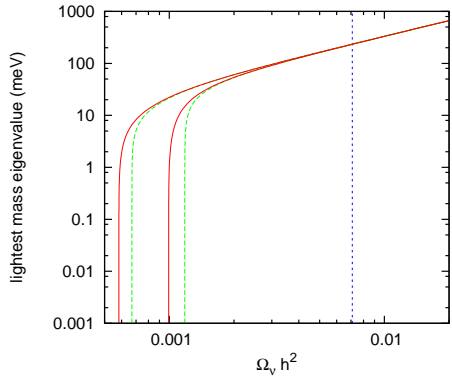}
\caption{\label{fig:momeganu} The lightest mass eigenvalue as a function of $\Omega_\nu h^2$.  Solid curves use the quantum mechanical calculation method described here, while dashed curves use the naive method illustrated in Eq. (\ref{eq:rhonu_naive}).  Curves on the left correspond to the normal neutrino mass hierarchy; curves on the right are for the inverted hierarchy.  Also plotted is the limit from the WMAP 5-year data, $\Omega_\nu h^2 < 0.0071$ ($95 \%$ confidence limit).}
\end{figure}

This is a concern since the naive computation of neutrino energy densities differs from the quantum mechanical method.  Fig.\ \ref{fig:momeganu} shows the relationship between the lightest mass eigenvalue ($m_1$ in the normal hierarchy or $m_3$ in the inverted hierarchy) and the measured value of $\Omega_\nu$.  We see that the limit on $\Omega_\nu h^2$ for excluding the inverted hierarchy is reduced by about $15 \%$. (Here $h$ is the Hubble parameter in units of $100 ~\rm{km} ~\rm{s}^{-1} ~\rm{Mpc}^{-1}$.)  In addition, at low mass values, we see that the mass is a very sensitive function of $\Omega_\nu h^2$.  As a result, mass measurements using this method are sensitive to the precision in the inferred measurement of $\Omega_\nu h^2$.  A $10 \%$ error in the measurement of $\Omega_\nu h^2$ would prevent us from inferring a lightest neutrino mass $\lesssim 5 ~\rm{meV}$.  
For neutrino masses $\lesssim 10 - 20 ~\rm{meV}$, there can be a significant error in determining the lightest mass eigenvalue if the quantum mechanical treatment we present here was not used.


The current observational limits (shown in 
Fig.\ \ref{fig:momeganu}
as the five-year WMAP data \cite{wmap5}), are nowhere near the regime where a consistent quantum mechanical treatment is necessary.  However, the quantum mechanical analysis may be relevant for higher precision future neutrino mass inferences based on transfer function arguments.

One possible complication has not been discussed.  It has been claimed that the phase space density of the relic neutrinos is high so that quantum coherence between momentum states may develop as a result of fermion exchange symmetry \cite{pm06}.  We note that the initial distribution of neutrinos, Eq.\ (\ref{dist}), is determined by Fermi-Dirac quantum statistics because it reflects emergence from an environment where thermal equilibrium obtains.  Moreover, the general relativistic analog to Liouville's Theorem assures us that as these neutrinos freely stream through spacetime, their proper phase space density remains unchanged and therefore continues to respect the Pauli principle.  

\begin{figure}
\includegraphics[width = 2in, angle = 270]{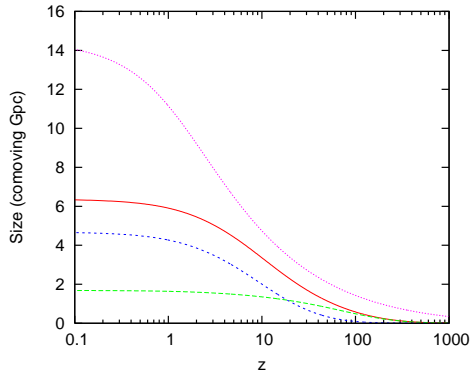}
\caption{\label{fig:wp} The separation of mass eigenstates as a function of redshift.  Solid curve for $\nu_1 - \nu_3$, long dashed curve for $\nu_2 - \nu_3$, and medium dashed curve for $\nu_1 - \nu_2$.  The causal horizon is the short dashed curve.}
\end{figure}

There is a second, peculiar effect arising from the fact that these relic neutrinos decouple in flavor eigenstates.  A decoupled relic neutrino $\nu_\alpha$ can be regarded as a coherent superposition of mass states with common spacelike momentum ${\bf p}$. As the universe expands and this momentum redshifts downward, there will come an epoch where the higher rest mass component of this flavor wavepacket will tend toward a nonrelativistic velocity, while the lower mass components speed along at nearly the speed of light. From this point onward the spatial scale of the coherent flavor wavepacket will grow rapidly with time. The net result is that at the present epoch, and at the redshift $z \sim 1$ epoch where neutrinos are falling into dark matter potential wells, the relic neutrinos are coherent structures with sizes considerably larger than the spatial scale of the gravitational potential wells of galaxies and, in some cases, comparable to the causal horizon size. 
Using the normal neutrino mass hierarchy scheme and $m_1 = 1 ~\rm{meV}$, Fig.~\ref{fig:wp} shows how the size of a relic neutrino flavor wavepacket would grow with the expansion of the universe. 



The more massive components of these neutrino flavor wave packets become nonrelativistic very early on, and it is interesting that the two different possibilities for the neutrino mass hierarchy yield quite different kinematic histories for the relic neutrinos. For example, with $m_1 =1\,{\rm meV}$ in the normal neutrino mass hierarchy, the measured neutrino mass-squared differences imply that $m_2 \approx 8.7\,{\rm meV}$, and $m_3 \approx 49\,{\rm meV}$.  As the wave packets for the three mass eigenstates redshift in the expanding universe, the components corresponding to each of the three mass eigenstates tend to become non-relativistic at three distinct epochs, $z \sim 5$, $50$, and $300$ respectively.  

On the other hand, the inverted neutrino mass hierarchy would yield a different history.  The three neutrino masses would be $m_3 = 1 ~\rm{meV}$, $m_1 \approx 49 ~\rm{meV}$, and $m_2 \approx 50 ~\rm{meV}$, which would imply that the lightest mass eigenstate would tend to be non-relativistic at $z \sim 5$ while the two heavier mass eigenstates would become non-relativistic at $z \sim 300$.

If the lightest neutrino mass eigenvalue is $ \ll 1\,{\rm meV}$, then the relic neutrinos might be coherent flavor eigenstates consisting of superpositions of relativistic and nonrelativistic components at the epoch $z\approx 2$ when these particles begin to fall into the gravitational potential wells associated with galaxies. Though the role of spacetime curvature in quantum state reduction is not settled \cite{pen96, chr05,fey95}, an obvious set of questions is posed. 

First, does the process of ``capturing'' a neutrino into a gravitational potential well lead to flavor wavepacket de-coherence? Given the disparity between the relatively small local spacetime curvature scale and the flavor wavepacket size in this example, it is plausible that the gravitational tidal stress induces de-coherence. In this picture, the neutrino clustering/capture process would be tantamount to a ``measurement,'' with capture occurring when flavor wave function collapse yields a nonrelativistic mass-energy eigenstate. 

However, this line of reasoning calls into question the premise of coherent flavor eigenstates at the epoch when neutrinos are captured into, {\it e.g.,} clusters of galaxies. Fig.\ \ref{fig:wp} shows that the relic neutrino flavor wavepackets can have spatial extents that are comparable to the spacetime curvature scale of the universe itself.   
In this picture, the complete de-coherence history of the relic neutrinos could be obtained only through solution of a fully covariant Dirac-like field development equation with all matter and dark matter distributions {\it plus} a prescription for de-coherence (wave function collapse). 
An alternative possibility is that de-coherence would never happen. 

In either case, unitarity implies that the comoving number density of neutrinos at a given mass is fixed.  As a result, unless there is new physics associated with curvature-induced decoherence, the accretion history of neutrinos in potential wells should depend only on the neutrino masses and, in particular, the mass hierarchy.

The on-going revolutions in observational cosmology and experimental neutrino physics overlap when it comes to the relic neutrinos, among the most abundant particles in the universe.  The fruits of this symbiotic overlap include arguably the best probe of neutrino masses.  The work presented here directly addresses these neutrino mass issues, but also presents new insights into the role of quantum coherence and decoherence in the history of these relic particles.


This work was supported in part by
NSF grant PHY-06-53626.
We thank B.~Keating, Y.~Raphaelli, E.~Gawiser and M.~Shimon
for valuable conversations.

\bibliography{wavepacket}

\end{document}